\documentclass[a4,aps,amsmath,floatfix,nofootinbib,superscriptaddress,10pt]
{revtex4}
\usepackage{graphicx} \usepackage{color} \usepackage{enumerate} 
\newcommand{\be}{\begin{equation}} \newcommand{\ee}{\end{equation}} 
\newcommand{\ba}{\begin{array}} \newcommand{\ea}{\end{array}} 
\newcommand{\bea}{\begin{eqnarray}} \newcommand{\eea}{\end{eqnarray}} 
\newcommand{\bdm}{\begin{displaymath}} 
\newcommand{\edm}{\end{displaymath}}

\begin{document}

\title{Critical hysteresis on dilute triangular lattice}

\author{Diana Thongjaomayum}
\affiliation{Center for Theoretical Physics of Complex Systems,\\
Institute for Basic Science (IBS), Daejeon 34051, Republic of Korea}
\author{Prabodh Shukla}
\affiliation{North-Eastern Hill University, Shillong-793022, India}
\thanks{Affiliation at retirement}

\begin{abstract}

{Critical hysteresis in the zero-temperature random-field 
Ising model on a two-dimensional triangular lattice has been studied 
earlier with site dilution on one sublattice. It was reported that 
criticality vanishes if less than one third of the sublattice is 
occupied. This appears at variance with recently obtained exact 
solutions of the model on dilute Bethe lattices and prompts us to 
revisit the problem using an alternate numerical method. Contrary to 
our speculation that criticality may not be exactly zero below one 
third dilution, the present study indicates it is nearly zero if 
approximately less than two-thirds of the sublattice is occupied. This 
suggests that hysteresis on dilute periodic lattices is qualitatively 
different from that on dilute Bethe lattices. Possible reasons are 
discussed briefly.}

\end{abstract}

\maketitle

\section{Introduction}

The random-field Ising model~\cite{imryma} was introduced to study the 
effect of quenched disorder on a system's ability to sustain long-range 
order in thermal equilibrium. After a rather prolonged debate it was 
resolved that the lower critical dimension of the Ising 
model~\cite{ising} remains equal to two in the presence of quenched 
random fields~\cite{imbrie}. Subsequently a zero-temperature version of 
the model (ZTRFIM) without thermal fluctuations but an on-site quenched 
random field distribution N[0,$\sigma^2$] was introduced 
~\cite{sethna1, sethna2} as a model for disorder-driven hysteresis in 
ferromagnets and other similar systems~\cite{sethna3}. Numerical 
simulations of ZTRFIM on a simple cubic lattice reveals a critical 
value of $\sigma = \sigma_c \approx 2.16 J$ ($J$ being the nearest 
neighbor ferromagnetic exchange interaction). For $\sigma < \sigma_c$ 
each half of the hysteresis loop shows a discontinuity in 
magnetization. The size of the discontinuity decreases to zero at a 
critical value of the applied field $h_c \approx 1.435 J$ as $\sigma$ 
is increased to $\sigma_c$. The behavior near $\{h_c,\sigma_c\}$ shows 
scaling and universality quite similar to the one caused by critical 
thermal fluctuations at an equilibrium critical point. These aspects of 
the model are important in understanding hysteresis experiments and 
related theoretical issues. Initial numerical attempts to find a 
$\sigma_c$ on the square lattice were inconclusive casting doubt on the 
lower critical dimension of the model. More extensive simulations 
~\cite{spasojevic} indicate $\sigma_c \approx 0.54 J$ and $h_c \approx 
1.275 J$ on the $2d$ square lattice.

An exact solution~\cite{dhar} of ZTRFIM on a Bethe lattice of integer 
connectivity $z$ shows that criticality occurs only if $z \ge 4$. 
Normally critical behavior on Bethe lattices is independent of $z$ if 
$z>2$ and is the same as in the mean field theory. Therefore the result 
for hysteresis is unusual and efforts have been made ~\cite{handford} 
to understand the physics behind it. A useful insight is obtained by 
extending the analysis to noninteger values of $z$ ~\cite{shukla1, 
shukla2}. This is done by considering lattices where the connectivity 
of each node is distributed over a set of integers so that the average 
connectivity of a node has a noninteger value greater than two. 
Fortunately the problem can still be solved exactly and leads to the 
identification of a general criterion for the occurrence of critical 
hysteresis. The general criterion is that there should be a spanning 
path across the lattice and a fraction of sites on this path (even an 
arbitrarily small fraction) should have connectivity $z \ge 4$.

On periodic lattices, an exact solution of ZTRFIM is not available.  
Extant simulations indicate that the existence or absence of $\sigma_c$ 
on a periodic lattice with uniform connectivity $z$ is the same as on a 
Bethe lattice of connectivity $z$. Criticality is absent on any lattice 
with $z=3$ irrespective of the dimension $d$ of the space in which the 
lattice is embedded ~\cite{sabhapandit}. Indeed critical hysteresis 
appears to be determined by a lower integer connectivity $z_\ell=4$ 
rather than a lower critical dimension $d_\ell=2$. As $z$ increases 
above $z_\ell$, the critical point becomes easier to observe in 
simulations. Compared with the intensive simulations on large square 
lattices, it takes a modest effort to observe criticality on a 
triangular lattice ~\cite{diana}. However the estimated value of 
$\sigma_c$ appears to decrease slowly with increasing size of lattice. A 
study on lattices of size $L \times L$ with $L \le 600$ gives 
$\sigma_c=1.22$ ~\cite{kurbah}, while more extensive simulations on 
lattices of size up to $L \le 65536$ yield $\sigma_c=0.85$ 
~\cite{janicevic}. We may remark that the critical exponents on the 
triangular lattice appear to be different from those on the square 
lattice ~\cite{janicevic}. This is puzzling in the context of the 
universality of critical phenomena and the broader implications of this 
result are not clear. At present $L=65536$ is the largest linear size 
that has been studied thoroughly using available computers. One may ask 
if $\sigma_c$ would decrease further in case much larger values of $L$ 
were studied. Although extant numerical studies do not suggest $\sigma_c 
\to 0$ as $L \to \infty$ but we are not aware of a rigorous argument for 
the same. Questions of this nature can not be resolved conclusively by 
numerical studies. Criticality on a dilute lattice is even harder to 
settle numerically due to additional positional disorder. Keeping this 
in mind, our focus in the present paper is on systems of modest sizes 
and try to understand the qualitative trends in the basic data.

It has been argued that $\sigma_c=0$ for an asymmetric distribution of 
the random field in case $z=3$ and $\sigma_c > 0$ for integer values $z 
\ge 4$ ~\cite{sabhapandit}. Our object here is to examine non-integer 
values of $z > 3$. A dilute (partially occupied) lattice of 
connectivity $z$ enables us to study a lattice of average connectivity 
$z_{av} < z$. We consider a triangular lattice $T=A+B+C$  with one 
of its constituent sublattices, say $C$, having a reduced occupation 
probability $c$ ~\cite{kurbah}. The average connectivity on $T$ is then 
equal to $z_{av}=6(1+2c)/(2+c)$ and the average connectivity on $A$ or 
$B$ sublattice is equal to $z_{eff}=3(1+c)$. The connectivity of 
occupied sites on $C$ is equal to six. As $c$ is reduced from $1$ to 
$0$, we go from a triangular to a honeycomb lattice. Extant work 
indicates that $\sigma_c$ drops to zero at $c=1/3$ within numerical 
errors. At $c=1/3$, $z_{eff}=4$. Keeping in mind that $z 
\ge 4$ is required for criticality on lattices of uniform integer 
connectivity $z$, it does look reasonable at first sight that 
$\sigma_c=0$ for $c < 1/3$ on a diluted lattice. However recent studies 
~\cite{shukla1,shukla2} on Bethe lattices of mixed coordination number 
bring out a new twist in the importance of sites with connectivity $z 
\ge 4$. Criticality has been shown to exist even if a fraction of 
occupied sites have $z < 4$ but there should be a spanning path through 
occupied sites and a fraction of sites on this path should have $z \ge 
4$. If this criterion were to apply to dilute periodic lattices as 
well, we may expect a non-zero $\sigma_c$ in the entire range $1 \ge c
> 0$.

The reason for a discontinuity in the hysteresis loop on a Bethe lattice 
is that a fixed point corresponding to zero magnetization becomes 
unstable and splits into two stable fixed points for $\sigma < 
\sigma_c$. The size of the splitting is the size of the discontinuity. 
This is easily demonstrated by an analysis of the model on a Cayley tree 
~\cite{shukla2}. We set the applied field equal to zero, and consider an 
initial configuration with all spins down except the spins on the 
surface of the tree. If the surface spins are equally likely to be up or 
down i.e. if the surface magnetization is zero it remains zero as spins 
are relaxed layer by layer towards the interior of the tree. Small 
perturbations to the surface magnetization behave differently depending 
on the connectivity $z$ of the lattice. If $z \le 3$ the perturbations 
decrease and the magnetization in the deep interior remains zero. If $z 
\ge 4$, the perturbations diverge. A positive value of magnetization 
tends to increase, and and a negative value tends to decrease as we move 
towards the interior. An important point is that this is not just a 
global property of a lattice of uniform connectivity $z$. On a lattice 
with mixed connectivity, each node depending on its connectivity $z$ 
increases or decreases the perturbation passing through it in a similar 
fashion. Larger the connectivity of the node, larger is the enhancement. 
Thus a small perturbation on the surface leads to a finite discontinuity 
in the deep interior of the tree if a fraction of nodes along the path 
have $z \ge 4$. Of course a spanning path is a prerequisite to reach the 
deep interior. However spanning paths are always there under our scheme 
of dilution. Even if $c=0$, there are spanning paths on the honeycomb 
lattice; $c>0$ introduces additional paths containing $C$ sites. The $C$ 
sites have connectivity equal to six. As long as there are some $C$ 
sites there are spanning paths punctuated by sites with connectivity 
equal to six. Remaining sites, the $A$ and $B$ sites have connectivity 
$3(1+c)$ on the average. If $c \ge 1/3$ the average connectivity of each 
site on the spanning path is greater than four and we have a case for a 
relatively large discontinuity as observed in extant simulations. On the 
other hand, if $c < 1/3$, we should still expect a discontinuity albeit 
a much smaller one. The argument in favor of it is the enhancement 
effect of nodes with $z\ge4$ on a Bethe lattice. It is not clear {\it{a 
priori}} how loops on a periodic lattice may vacate this effect. This 
forms the motivation to review critical hysteresis on the dilute 
triangular lattice. However simulations presented below suggest that 
criticality on a dilute triangular lattice is qualitatively different 
from that on a dilute Bethe lattice.

It may not be out of place to make two general remarks on hysteresis 
studies in ZTRFIM before getting into the specifics of the present 
paper. Firstly, setting temperature and driving frequency equal to zero 
is an approximation. Hysteresis in physical systems is necessarily a 
finite temperature and finite time phenomena. A key feature of ZTRFIM 
is the occurrence of a fixed point under the zero-temperature dynamics. 
Scale invariance around the fixed point is directly related to 
experimental aspects of Barkhausen noise. The fixed point at $\sigma_c$ 
is lost if any of the two approximations are relaxed ~\cite{shukla3}. 
This is disconcerting but does not end the usefulness of ZTRFIM. The 
model has been applied to a variety of social phenomena including 
opinion dynamics where the zero temperature Glauber dynamics is not so 
unrealistic ~\cite{shukla4}. Therefore efforts to improve our technical 
understanding of ZTRFIM on different lattices and their associated 
universality classes would remain of value in statistical mechanics.

\newpage

\section{The Model and Numerical Results}

In order to make the paper self-contained and better 
readable, we describe the model briefly. Readers may refer to 
~\cite{kurbah} for more details. The Hamiltonian is, \be 
H=-J\sum_{i,j}s_is_j -\sum_{i}h_is_i-h\sum_{i}s_i. \ee J is 
ferromagnetic interaction, the double sum is over nearest neighbors of 
a 2d triangular lattice of size $N=L \times L$; $s_i=\pm1, 
i=1,\ldots,N$ are Ising spins; $h_i$ is a quenched random field drawn 
from a distribution N(0,$\sigma^2$) and $h$ is an external field that 
is ramped up adiabatically from $-\infty$ to $\infty$ and back down to 
$-\infty$. The triangular lattice comprises three sublattices $A$, $B$, 
and $C$; $A$ and $B$ are fully occupied but sites on $C$ sublattice are 
occupied with probability $c$ $(0\le c \le 1)$. Thus we have a 
triangular lattice at $c=1$, but a honeycomb lattice at $c=0$. 
Hysteresis under zero-temperature Glauber dynamics is studied as 
follows. Depending upon the size of the system N, we start with a 
sufficiently large and negative $h=-h_0$ such that the state 
$\{s_i=-1\}$ is stable. A stable configuration has each spin $s_i$ 
aligned along the local field $\ell_i= n J -(z-n) J + h_i + h $ at its 
site; here $z$ is the number of nearest occupied neighbors of $i$; $n$ 
neighbors being up ($s=1$), and $(z-n)$ down ($s=-1$). The 
magnetization per spin in a stable state is $m(h)=N^{-1}\sum_i s_i$. 
Thus we start with a stable state with $m(-h_0)=-1$. Now we increase 
$h$ by the minimal amount, say $h=h_1=-h_0+\delta h_1$ that makes one 
of the spins unstable. An attempt to stabilize this spin may make some 
or all of its neighbors unstable. We hold $h=h_1$ constant and 
iteratively flip up unstable spins until no spins in the system are 
unstable. This results in an avalanche of flipped spins in the vicinity 
of the initial unstable spin. The increase of magnetization from 
$h=h_0$ to $h=h_1$ is equal to twice the size of the avalanche. Holding 
the applied field constant during the avalanche corresponds to the 
assumption that the applied field varies infinitely slowly in 
comparison with the spin relaxation rate. The stable state at the end 
of an avalanche corresponds to a local minimum in the energy landscape, 
and depends on the history of the system. In our example the local 
minimum retains memory of the initial state with $m(-h_0)=-1$. Under 
finite temperature Glauber dynamics, the system may escape the local 
minimum and move towards the global minimum albeit very slowly. For 
this reason we may occasionally refer to the stable state under zero 
temperature dynamics as a metastable state. Employing the above 
procedure repeatedly, we determine all the metastable states between 
$m(-h_0)=-1$ and $m(h_0)=1$ on lower half of the hysteresis loop, and 
similarly on the upper half as well. Fig.1 depicts the result for 
$c=0.90$ and $\sigma=0.9$ and $\sigma=2.5$ respectively. The key point 
is that for smaller $\sigma$ the loop has discontinuities while there 
is no discontinuity for larger $\sigma$. The upper and lower halves of 
the loop are related by symmetry and therefore it suffices to focus 
only on the lower half. Apparently, there is a critical value 
$\sigma=\sigma_c$ which separates discontinuity at $\sigma < \sigma_c$ 
from no discontinuity at $\sigma > \sigma_c$ but the numerical 
determination of $\sigma_c$ is a challenging task.

Our main interest is to understand the qualitative dependence of 
$\sigma_c$ on $L$ and $c$, and to check in particular if $\sigma_c$ 
drops to zero abruptly when $c$ drops below $c = 1/3$. The defining 
feature of $\sigma_c$ is that the discontinuity in the magnetization 
$m(h)$, say on the lower half of the hysteresis loop, reduces to zero as 
$h \to h_c$ and $\sigma \to \sigma_c$ from below. Exact solution on 
Bethe lattice and simulations on periodic lattices reveal that a 
discontinuity in magnetization is accompanied by a reversal of 
magnetization. Numerical determination of a discontinuity is rather 
problematic. For small $\sigma$, the graph $m(h)$ vs. $h$ near $m(h)=0$ 
tends to be almost vertical anyway. A simulation based on a single 
realization of the random-field distribution necessarily shows a broken 
curve comprising a few irregularly placed discontinuities due to large 
fluctuations in the system. The number as well as positions of 
discontinuities vary from configuration to configuration and averaging 
over configurations results in a steep but smooth $m(h)$ curve. A 
genuine underlying discontinuity, if any, has to be inferred from the 
character of fluctuations. An added complication is that fluctuations at 
a discontinuity are different from those at the critical point where the 
discontinuity vanishes. Finally finite size scaling has to be employed 
to infer $\sigma_c$ in the thermodynamic limit. The estimate for 
$\sigma_c$ using finite size scaling should be independent of system 
sizes used in numerical simulations. However numerical uncertainties are 
large and diminish extremely slowly with increasing system size. As 
mentioned earlier, initial studies on triangular lattices of sizes $L 
\times L$ with $L \le 600$ indicated $\sigma_c=1.22$ ~\cite{kurbah} but 
more extensive simulations on lattices up to $L=65536$ yield $\sigma_c = 
0.85$ ~\cite{janicevic}. The procedure for determining $\sigma_c$ is 
rather indirect, tedious, cpu intensive, and various compromises have to 
be made in order to draw reasonable conclusions~\cite{kurbah}.

In this paper we adopt a different approach than used in previous 
studies. The basic idea is simple although the details have similar 
issues as in earlier studies. The new approach is useful in discerning 
important trends in the behavior of the model based on simulations of 
systems of modest sizes. For a fixed $\sigma$ on an $L\times L$ 
lattice, we count the total number of metastable states $M(\sigma;L)$ 
(fixed points under zero temperature Glauber dynamics) comprising the 
lower half of hysteresis loop. As indicated in the previous paragraph we 
increase the applied field $h$ by a minimal amount to go from one fixed 
point to the next and keep $h$ fixed during the relaxation process. We 
plot $M(\sigma;L)$ as a function of $\sigma$. It is a monotonically 
increasing function of $\sigma$ without any discontinuity. The cpu time 
increases rapidly with increasing $L$ and $\sigma$. Fig.2 shows the 
result on a modest $33 \times 33$ triangular lattice and $0 < \sigma 
\le 50$. The general features of Fig.2 are easy to understand. In the 
limit of small $\sigma$, $\sigma \le 0.4 $ approximately, the first 
spin to flip up initiates an infinite avalanche of flipped spins giving 
$M(\sigma < 0.4;L)=2$. In the limit $\sigma \to \infty$, spins flip up 
independently and $M(\sigma \to \infty;L)$ increases towards $L \times 
L$. We expect $M(\sigma;L)$ to increase continuously from $2$ to $L 
\times L$ as $\sigma$ increases from $0$ to $\infty$ on a finite 
lattice. This expectation is born out by Fig.2. If there is a critical 
value of $\sigma$ separating discontinuous $m(h)$ for $\sigma < 
\sigma_c(L)$ with continuous $m(h)$ for $\sigma > \sigma_c(L)$ we ought 
to see its signature in the $M(\sigma;L)$ graph. A discontinuity in 
$m(h)$ for $\sigma < \sigma_c$ would effectively reduce $M(\sigma;L)$ 
in proportion to its size. This would result in some change in shape of 
$M(\sigma;L)$ $ vs. $ $\sigma$ graph at $\sigma_c(L)$. We find that 
this effect is present but too weak to be seen with naked eye in the 
main graph of Fig.2 or its magnified portion in the range $0 < \sigma < 
4$ shown in the left inset there. However, we do see an apparent 
inflexion point around $\sigma \approx 1.8$ if $M(\sigma;L)$ is plotted 
on logscale scale as in the right inset.We tentatively 
identify this inflexion point with $\sigma_c(L)$, the critical 
$\sigma_c$ on an $L\times L$ lattice. A scaling property of 
$M(\sigma;L)$ with respect to $L$ presented below confirms this 
identification.

Fig.3 shows $\log_{10} M(\sigma;L)$ $ vs. $ $\sigma$ on a triangular lattice 
for $0 < \sigma <2$ and $L=33, 99, 198, 333, 666, 999$. The results 
have been averaged over $10^4$ configurations of the random field 
distribution for $L \le 198$ and $10^3$ configurations for $L \ge 333$. 
As expected, the graphs start at $\log_{10} 2$ and fan out towards $\log_{10} L$ 
with increasing $\sigma$. There is an apparent scaling with respect to 
$L$. Fig.4 brings out this scaling explicitly by plotting $G(\sigma;L)$ 
where $G(\sigma;L) = \log_{10}{ \frac{M(\sigma)}{L \times L}}$. The quantity 
$G(\sigma;L)$ is the logarithm of the density of metastable states per 
unit area of the lattice. Each $G(\sigma;L)$ has an apparent inflexion 
point at $\sigma_c(L)$ being concave up for $\sigma < \sigma_c(L)$, and 
convex up $\sigma > \sigma_c(L)$. Graphs for $\sigma > \sigma_c(L)$ 
merge into each other from above meaning they maintain their relative 
order in $L$ as they merge. It is easy to understand this behavior. 
Each metastable state is associated with an avalanche that precedes it. 
Therefore we may visualize a metastable state as an area on the $L 
\times L$ lattice occupied by spins that turn up together in an 
avalanche. It helps to understand the following discussion if we 
imagine coloring the area occupied by each avalanche with a different 
color. $G(\sigma,L)$ is then the logarithm of the density of colors 
when colors fill the entire lattice. Inverse of the density gives the 
average area occupied by a randomly chosen color. Independence of 
$G(\sigma;L)$ from $L$ for $\sigma > \sigma_c(L)$ suggests that colors 
are well dispersed and each color is spread over a much smaller area 
than $L\times L$. In other words, it suggests the absence of a large 
spanning avalanche of the order of $L\times L$. The curve is convex up 
because $G(\sigma;L)$ increases with increasing $\sigma$ and approaches 
saturation in the limit $\sigma \to \infty$. In contrast, $G(\sigma;L)$ 
for $\sigma < \sigma_c(L)$ depends on $L$ and is concave up. This too 
is understandable. In this regime, there is a spanning cluster on the 
scale $L\times L$. Let us color it black. The black cluster contributes 
merely one color to the lattice but takes up a disproportionately huge 
area preventing more colors from getting in. This significantly reduces 
$G(\sigma;L)$. The black cluster shrinks to zero as $\sigma \to 
\sigma_c(L)$ from below. The area vacated by the shrinking cluster is 
gradually filled up by smaller clusters of different colors thus 
increasing $G(\sigma;L)$. This explains the concave up shape as well as 
the $L$-dependence of $G(\sigma;L)$ for $\sigma < \sigma_c(L)$. These 
considerations lead us to associate the inflexion point on 
$G(\sigma;L)$ curve with $\sigma_c(L)$. In the following, we examine 
how $\sigma_c(L)$ shifts to lower values with increasing $L$. However, 
before describing the numerical work, we may draw attention to a 
practical limitation of our analysis.

We evaluate $G(\sigma)$ on six lattices of size $33 \le L \le 999$ for 
$0.1 J \le \sigma \le 2.0 J$. The range of $\sigma$ is chosen because 
$\sigma_c(L) \approx 1.8 J$ for $L=33$ and it is expected to decrease 
for larger $L$. We increment $\sigma$ in steps of $\delta \sigma =0.1$, 
getting $20$ data points for each $L$. Fitting the 20 points to a 
polynomial of degree 10 or so results in a reasonably good looking fit 
but the fitted curve has a wavy nature on a magnified scale. Taking the 
second derivative of the curve to find the inflexion point 
$\sigma_c(L)$ introduces errors and creates spurious inflexion points 
as well. To avoid the spurious inflexion points we adopt an alternate 
method which does not require fitting the data to a polynomial and 
serves to double check our results. We take $\sigma_c(L)$ as the point 
where $G(\sigma;L)$ vs $\sigma$ curve merges with the corresponding 
curve for the next higher value of $L$. In other words we take 
$G(\sigma;L)$ curve for $L=999$ as the boundary and $\sigma_c(L)$ for 
$L < 999$ as the point where the corresponding curve merges with the 
boundary. This procedure necessarily introduces an error due to the 
fixed increment $\delta \sigma$. In the absence of interpolations 
between values of $\sigma$ at fixed intervals, $\sigma_c(L)$ is restricted 
to one of the input values. However it produces qualitatively 
similar result as obtained by fitting the data to polynomials. We will 
return to this point when discussing our results in the following.

Let us call L33 the graph in Fig.4 corresponding to L=33 and similarly 
L99 etc. We find that L33 merges with L99 for $\sigma \ge 1.8$; L99 
merges with L198 for $\sigma \ge 1.5$; L198 merges with L333 for 
$\sigma \ge 1.4$; L333 merges with L666 for $\sigma \ge 1.3$; L666 
merges with L999 for $\sigma \ge 1.2$. As discussed in 
the preceding paragraph, we interpret these results as indicating 
$\sigma_c(L)=1.8,1.5,1.4,1.3,1.2$ for systems of linear size $L=33, 99, 
198, 333, 666$ respectively. If we fit $\sigma_c(L)$ to a power law 
scaling of the form \be \sigma_c(L)=\sigma_c + a \times L^{-b}\ee we 
find $\sigma_c(L)$ converges to $\sigma_c = 0.81\pm0.19$ in the limit 
$L\to \infty$ with $a= 2.85\pm0.37$ and $b= 0.30\pm0.09$. We have also 
fit $G(\sigma)$ $vs.$ $\sigma$ data to polynomials of degree eleven, 
and looked for inflexion points on the resulting continuous curve. 
Ignoring the spurious inflexion points near the boundaries of the range 
$[0.1 \le \sigma \le 2.0]$, we obtain $\sigma_c(L)=1.68, 1.48, 1.38, 
1.32, 1.25, 1.20$ for $L=33, 99, 198, 333, 666, 999$ respectively. 
Fitting these values to Eq.2 yields $\sigma_c=0.84\pm0.06$, 
$a=1.97\pm0.05$, and $b=0.24\pm0.03$. It is satisfying that the values 
of $\sigma_c$ obtained by the two methods are reasonably close to each 
other and also close to the estimate $\sigma_c=0.85\pm0.02$ obtained in 
reference ~\cite{janicevic} by studying large systems of size up to 
$L=65536$.

Simulations presented in Fig.4 demonstrate the existence of critical
hysteresis on a triangular lattice. Of course, the result is not 
new~\cite{diana, kurbah, janicevic}, but it validates a new method. Our 
goal is to apply the new method to examine criticality on dilute 
triangular lattice and compare with previous results ~\cite{kurbah}. In 
preparation for this goal we apply the new method to the case $c=0$ as 
well, i.e. on a honeycomb lattice. Fig.5 shows the results. 
Earlier studies have indicated the absence of critical 
hysteresis on a honeycomb lattice ~\cite{sabhapandit}. Therefore any 
prominent difference between the trends of Fig.4 and Fig.5 may be used 
as a tool to detect the presence or absence of critical hysteresis on a 
dilute lattice. Interestingly both figures have some common features as 
well as some prominent differences. Both show a threshold $\sigma_{th}$ 
such that $M(\sigma << \sigma_{th};L)=2$ and consequently $G(\sigma << 
\sigma_{th};L)= \log_{10}2 -2 \log_{10} L$. Thus in both cases the set of 
$G(\sigma;L)$ graphs for different $L$ are widely separated for $\sigma 
<< \sigma_{th}$ and merge into each other for $\sigma
>> \sigma_{th}$ as may be expected.

The prominent difference between Fig.4 and Fig.5 lies in the crossover 
from a set of widely separated curves at $\sigma << \sigma_{th}$ to 
their merger into each other at $\sigma >> \sigma_{th}$. On the 
triangular lattice, the curves maintain their relative order in $L$ but 
on the honeycomb lattice they reverse it. In the case $c=1$ each curve 
changes from concave up to convex up at the inflexion point 
$\sigma_c(L)$. As $L$ increases, $\sigma_c(L)$ decreases. In contrast, 
on the honeycomb lattice we do not see any clear indication of a 
inflexion point or a concave up portion. The threshold value of 
$\sigma$ below which $M(\sigma;L)=2$ depends on $L$ and varies somewhat 
from one configuration of random fields to another. The average over 
different configurations makes the curve rounded in this region but 
otherwise $G(\sigma;L)$ rises sharply with increasing $\sigma$ as well 
as increasing $L$. The sharp rise of $M(\sigma;L)$ with $\sigma$ and 
$L$ causes the reversal of the ordering of $G(\sigma;L)$ with respect 
to $L$ before the curves merge into each other from below.  This 
crossover takes place over a relatively narrow window $[0 ,\sigma]$ 
which shrinks further with increasing $L$ and moves towards lower 
$\sigma$. We take this to be a signature of the absence of criticality 
on finite lattices.
It is plausible that in the limit $L \to \infty$, the flat and concave up portions of the curves in Fig.5 may shrink to zero resulting in convex up curves over the full range $\sigma > 0$, but it is difficult to prove it 
numerically on lattice sizes studied here. The absence of critical 
hysteresis on a honeycomb lattice has been proven theoretically for an 
asymmetric distribution of the random field. It was shown if on-site 
quenched random fields are positive with the half-width of their 
distribution going to zero, $m(h)$ would increase smoothly from $-1$ to 
$1$ as $h$ increases from $-\infty$ to $J$. A similar argument can be 
used to prove that more than half spins in the system would have turned 
up continuously at $h=J$ for a Gaussian random field distribution. In 
other words, magnetization reversal would occur without a discontinuity 
as $\sigma \to 0$. Therefore critical hysteresis on the honeycomb 
lattice may be ruled out in the thermodynamic limit. Keeping in mind 
that finite size effects decrease logarithmically slowly, we take Fig.5 
as showing the absence of criticality on the honeycomb lattice.

The above discussion provides us a reasonable signature of critical 
hysteresis which can be read off from $G(\sigma;L)$ vs. $\sigma$ 
graphs. Fig.6, Fig.7, and Fig.8 show the results of simulations on 
dilute triangular lattice with $c=0.3, 0.4$, and $0.6$ respectively. It 
is evident that the case $c=0.3$ as well as $c=0.4$ is similar to the 
case $c=0$. Thus we conclude that critical hysteresis is absent in 
these cases. The graphs for $c=0.6$ seem to have a mixed character. 
Results for $L=33, 99, 198$ have the features of $c=0$ while those for 
$L=333, 666, 999$ appear closer in character to $c=1$. To us it seems 
that lattice with $c=0.6$ supports critical hysteresis albeit it is a 
borderline case. \\

\section{Discussion and Concluding Remarks}

Hysteresis in the zero-temperature random-field Ising model on 
honeycomb ($z=3$) and triangular ($z=6$) lattices is a difficult 
problem analytically as well as numerically.  Extant work indicates 
that honeycomb lattice does not support critical hysteresis but the 
triangular lattice does so. However the critical point $\sigma_c(L)$ on 
the triangular lattice decreases extremely slowly with increasing 
system size $L$. Intensive numerical simulations on large systems ($N 
\approx 10^{10}$) have been used to estimate $\sigma_c$ in the limit $L 
\to \infty$. The problem on the dilute triangular lattice is even more 
challenging. Simulations on modest systems ($N \approx 10^6$) along 
with finite size scaling and percolation arguments predict $\sigma_c=0$ 
for $c < 1/3$. At first sight this appears reasonable.  It is similar 
to the behavior on Bethe lattices of uniform integer connectivity; 
$\sigma_c > 0$ if $z \ge 4$. A dilute triangular lattice with $c < 1/3$ 
corresponds to $z_{av} < 4$ and one may expect it to have $\sigma_c = 
0$. However, recently obtained  exact solutions of the model on 
non-integer Bethe lattices predict $\sigma_c > 0$ for $3 < z_{av} \le 
4$. If similarity between Bethe and periodic lattices were to hold in 
general, it would mean $\sigma_c > 0$ on dilute triangular lattice for 
$0 < c < 1/3$ as well. The motivation for the present work was to 
examine this point.

We have used an approach based on the number of 
metastable states in the system $M(\sigma;L)$ and $G(\sigma;L) = \log_{10} 
M(\sigma;L) -2 \log_{10}(L)$. For a random field distribution 
N(0,$\sigma^2$) on lattices of size $33 \le L\le 999$ we find 
$G(\sigma;L)=\log_{10} 2 -2 \log_{10} L$ in the range $0 \le \sigma \le 0.3$. It  
rises monotonically towards zero in the limit $\sigma \to \infty$. The 
manner of rise depends on $c$ and indicates whether criticality is 
present or not. Drawing upon a general agreement in the literature that 
critical hysteresis exists for $c=1$ but not for $c=0$, we take the 
differences in the behavior of $G(\sigma;L)$ for these two cases as 
signatures of the presence or absence of criticality on a diluted 
lattice. The signatures are as follows. Consider $G(\sigma;L_1)$ and 
$G(\sigma;L_2)$ with $L_2 > L_1$. At very small values of $\sigma$, we 
have $G(\sigma;L_1) > G(\sigma;L_2)$. If criticality is present this 
order is maintained as both graphs go from concave up to convex up at 
inflexion points $\sigma_c(L_1)$ and $\sigma_c(L_2)$ respectively. For 
$\sigma \ge \sigma_c(L_1)$, $G(\sigma;L_1)$ merges with $G(\sigma;L_2)$ 
from above. If criticality is not present, the graphs do not show an 
inflexion point. Both appear convex up but $G(\sigma;L_2)$ overtakes 
$G(\sigma;L_1)$ before it merges with it from below for larger 
$\sigma$. These signatures are understandable consequences of the 
presence or otherwise of an infinite avalanche in the system. Apart 
from the absence of an infinite avalanche that causes $G(\sigma;L)$ to 
rise sharply with increasing $\sigma$, the connectivity $z$ of the 
lattice also plays a role. Lattices which do not support criticality 
have a lower connectivity e.g. $z=3$ for the honeycomb lattice. A 
typical avalanche on such lattice is smaller because there are lesser 
number of pathways going out from an unstable site to a potentially 
flippable site.

Somewhat unexpectedly the simulations presented here indicate 
$\sigma_c=0$ for $0 < c < 0.6$ approximately. We have used systems of 
the same order ($N \approx 10^6$) as used in ~\cite{kurbah} but 
processing of data under finite size scaling hypothesis has been 
avoided. The reason is that even if there is a theoretical argument for 
$\sigma_c \to 0$ as $L \to \infty$, a finite system would necessarily 
have an instability in the region $\sigma < \sigma_{th}$ where the first 
spin to flip up would cause all other spins to flip up as well. 
Fluctuations are extremely large in this region and finite-size scaling 
used in reference ~\cite{kurbah} may not be reliable. Fig.6-Fig.8 show 
that $\sigma_{th}$ is in the same ballpark as $\sigma_c$ predicted by 
finite size scaling in the range $0 \le c \le 0.6$. Thus an alternate 
method used in the present paper may be more reliable and a correction 
in earlier results is warranted. We note that earlier results 
~\cite{kurbah} also showed a change of behavior at $c \approx 0.6$. 
Table II and Fig.6 of reference ~\cite{kurbah} show a nearly linear 
decrease of $\sigma_c$ from $1.22$ at $c=1$ to $0.33$ at $c=0.6$; more 
rapid decrease to $0.26$ at $c=0.5$; then a constant value equal to 
$0.25$ at $c=0.40$ and $c=0.34$ before an abrupt drop to $0$ at 
$c=0.33$.

The change of behavior near $c=0.6$ and qualitative difference from 
dilute Bethe lattices most likely originates from closed loops on the 
diluted triangular lattice. It appears that closed loops on a lattice 
affect critical hysteresis more strongly than we expected beforehand. 
There are other indications as well. A square lattice is similar to a 
$z=4$ Bethe lattice in that both have the same connectivity and support 
critical hysteresis but $\sigma_c$ is quite different on the two 
lattices;$\sigma_c=0.54$ on a square lattice and $\sigma_c=1.78$ on a 
$z=4$ Bethe lattice. The difference is even more pronounced between a 
simple cubic and $z=6$ Bethe lattice. A diluted lattice has positional 
disorder as well as the random field. Although the average connectivity 
of the diluted lattice varies linearly with $c$ but the fraction of 
nodes with different connectivities vary differently with $c$. This 
possibly changes the nature of loops on the lattice. Our work 
suggests that positional disorder on a periodic lattice has a much 
stronger effect on $\sigma_c$ than it has on a Bethe lattice.

DT acknowledges the support from Institute for Basic Science in Korea 
(IBS-R024-D1).

\begin{figure}[ht] 
\includegraphics[width=0.8\textwidth,angle=0]{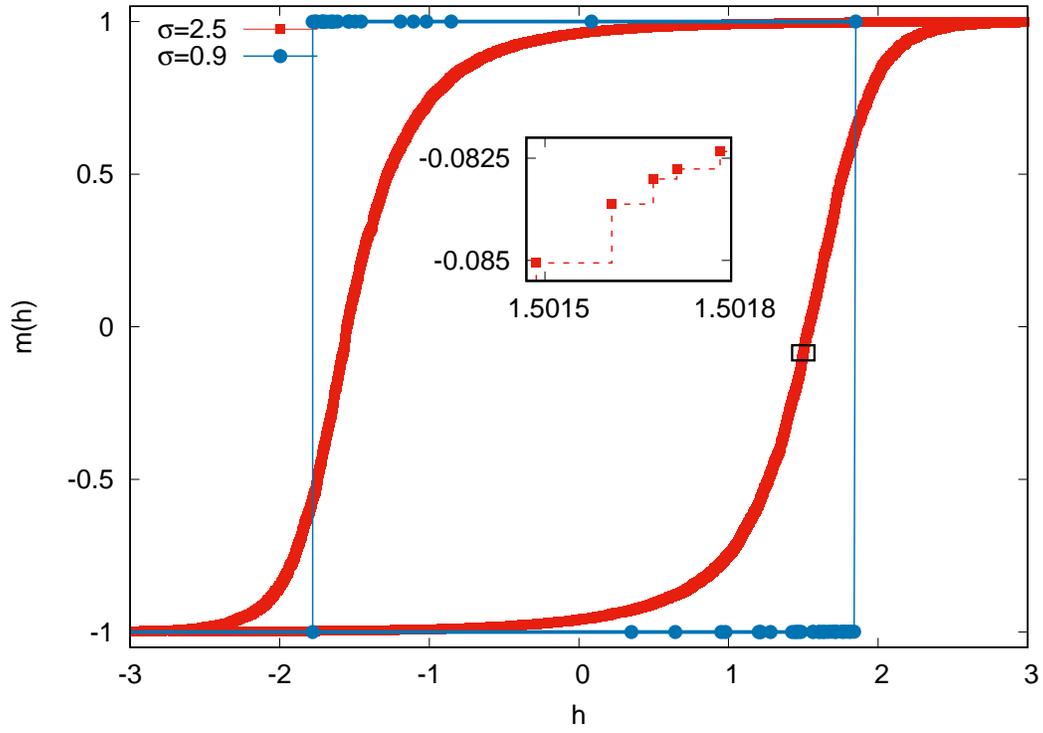} \caption{ 
Hysteresis loops in the zero-temperature random-field Ising model for 
N(0,$\sigma^2$) distribution of the random-field on an $L \times L$ 
triangular lattice with $L=333$ and $\sigma_c(L) \approx 1.2$. Loops are 
discontinuous for $\sigma < \sigma_c(L)$ but macroscopically smooth for 
$\sigma \ge \sigma_c(L)$. At microscopic level they exhibit Barkhausen 
noise as shown in the inset.}
\label{fig1} \end{figure}

\begin{figure}[ht] 
\includegraphics[width=0.8\textwidth,angle=0]{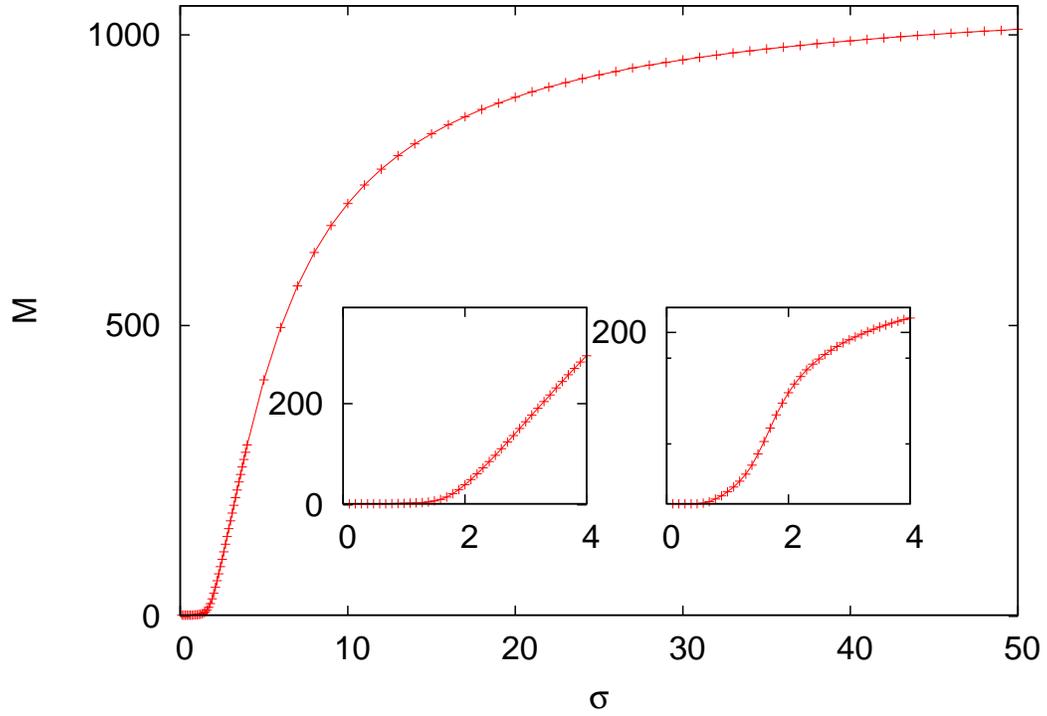} 
\caption{Number of metastable states M $vs.$ $\sigma$ comprising the 
lower half of hysteresis loop on a triangular lattice of size $33 
\times 33$ for Gaussian random field distributions N[0,$\sigma^2$]; for 
each $\sigma$ the value of M is averaged over $10^4$ independent 
realizations of the distribution. The first inset on the left shows an 
enlarged view of the graph in the range $0 < \sigma < 4$. The second 
inset on the right shows the same data as in the first inset but on a 
logarithmic scale along the y-axis.} \label{fig2} \end{figure}

\begin{figure}[ht] 
\includegraphics[width=0.8\textwidth,angle=0]{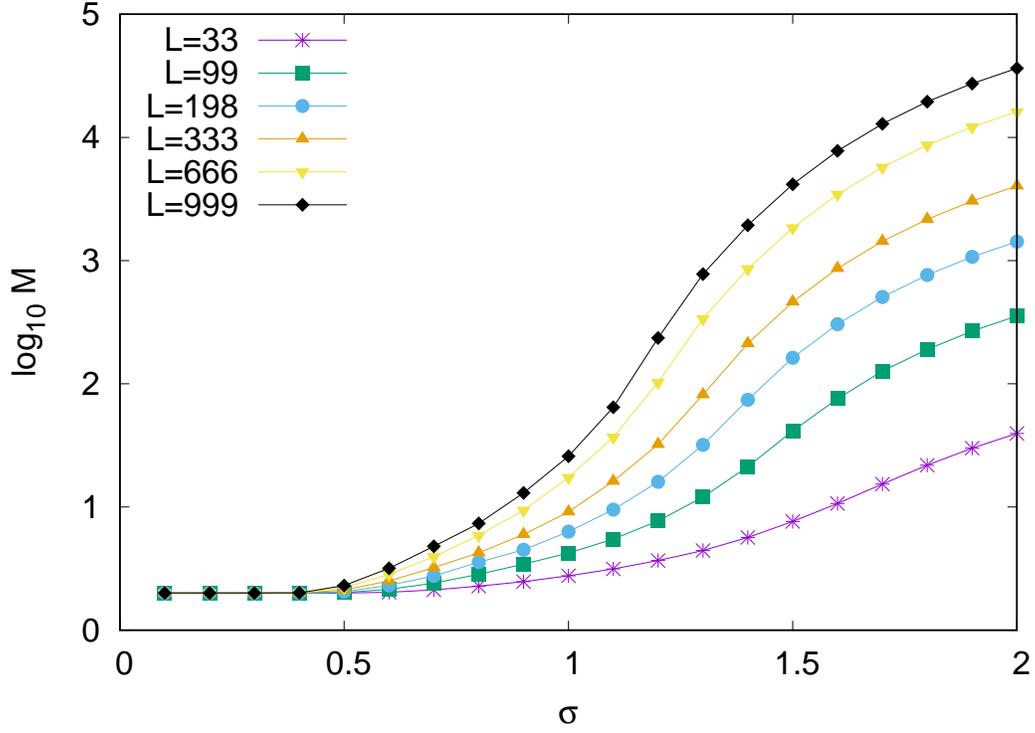} \caption{$\log_{10} 
M$ $vs.$ $\sigma$ in the range $0 < \sigma <2$ on an $L \times L$ 
triangular lattice; $L=33, 99, 198, 333, 666$, and $999$. M is averaged 
over $10^4$ configurations for $L \le 198$ and $10^3$ configurations 
for $L > 198$. As $L$ increases, the inflexion point $\sigma_c(L)$ in 
the corresponding graph shifts to lower $\sigma$.} \label{fig3} 
\end{figure}

\begin{figure}[ht] 
\includegraphics[width=0.8\textwidth,angle=0]{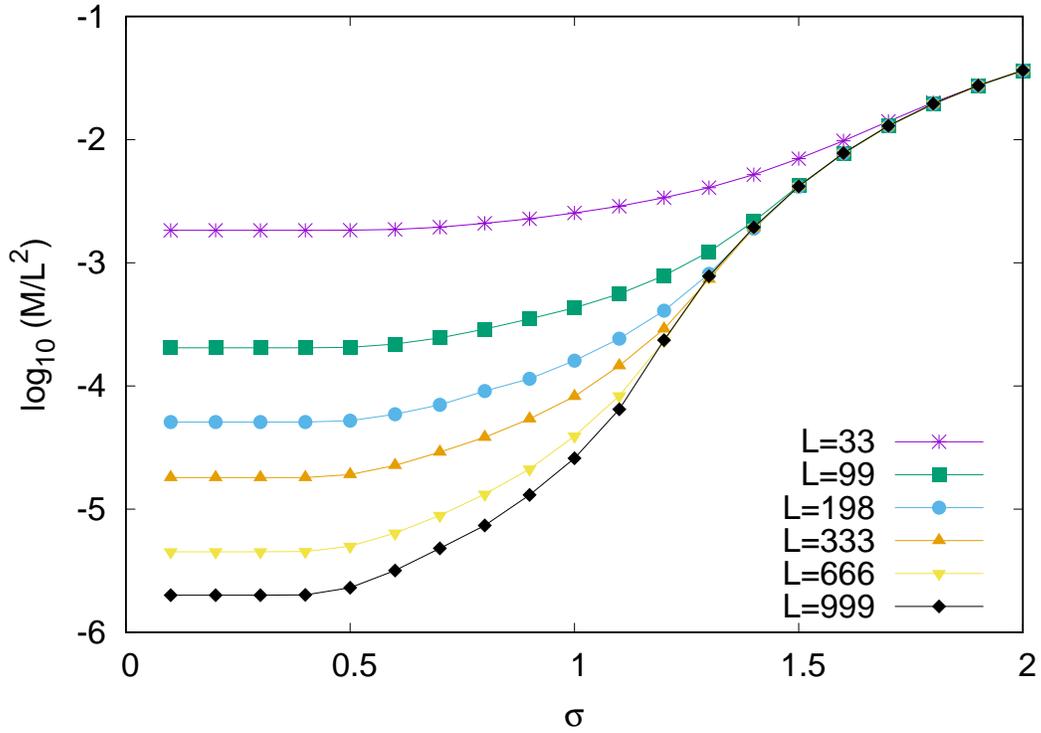} \caption{ The 
data in Fig.2 for the triangular lattice is replotted to show $\log_{10} 
\frac{M}{L^2}$ vs. $\sigma$. $M \to 2$ as $\sigma \to 0$ irrespective 
of $L$. It is therefore an artifact of scaling that graphs fan out for 
smaller values of $\sigma$. The interesting feature is that they 
overlap for larger values of $\sigma$. This feature may be exploited to 
estimate $\sigma_c$.} \label{fig4} \end{figure}

\begin{figure}[ht] \centering 
\includegraphics[width=0.8\textwidth,angle=0]{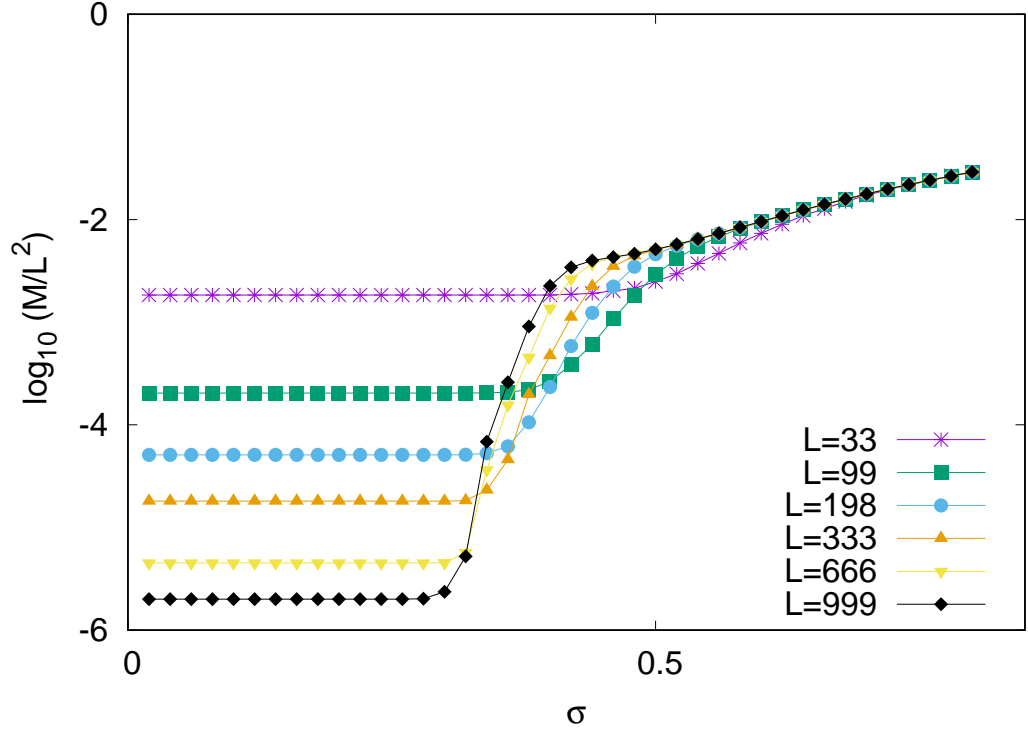} \caption{ $Log_{10}\frac{M}{L^2}$ vs. $\sigma$ on an $L \times L$ honeycomb lattice for 
$0 < \sigma \le 0.8$ and different values of $L$. A comparison with Fig.3 
indicates that critical hysteresis is absent on the honeycomb lattice 
(see text).}\label{fig5}\end{figure}

\begin{figure}[ht] \centering 
\includegraphics[width=0.8\textwidth,angle=0]{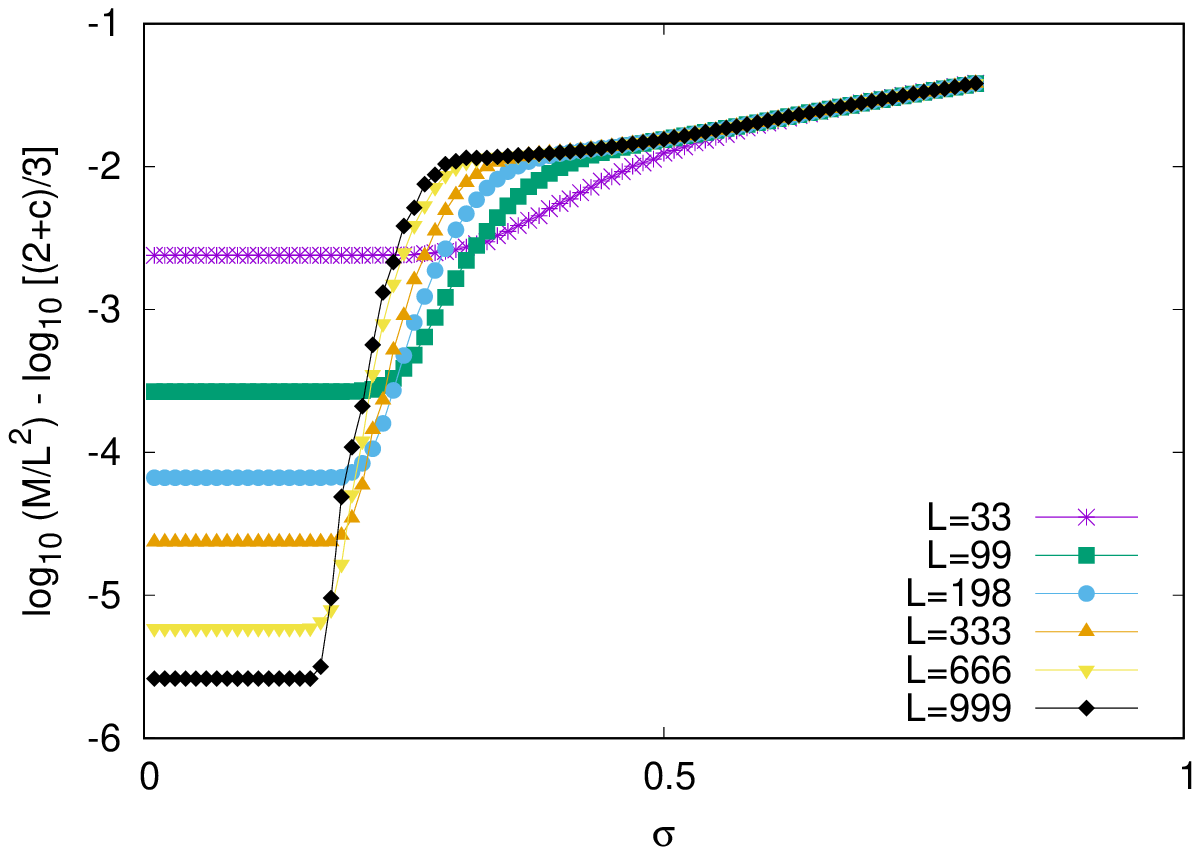} \caption{ 
Metastable states vs. $\sigma$ on a partially diluted triangular 
lattice with $c=0.30$ in the range $0.1 \le \sigma 
\le0.8$.}\label{fig6}\end{figure}

\begin{figure}[ht] \centering 
\includegraphics[width=0.8\textwidth,angle=0]{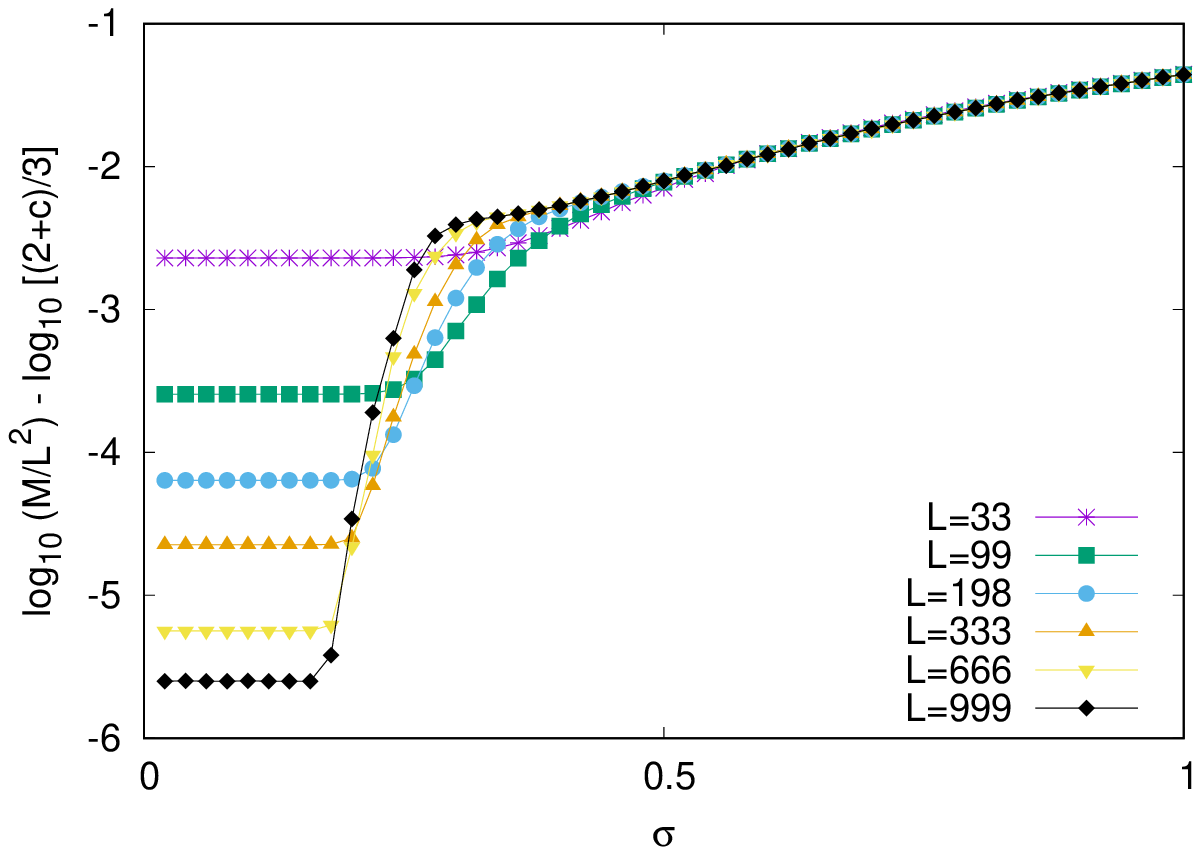} \caption{ 
Metastable states vs. $\sigma$ on a partially diluted triangular 
lattice with $c=0.40$ in the range $0.1 \le \sigma 
\le1.0$.}\label{fig7}\end{figure}

\begin{figure}[ht] \centering 
\includegraphics[width=0.8\textwidth,angle=0]{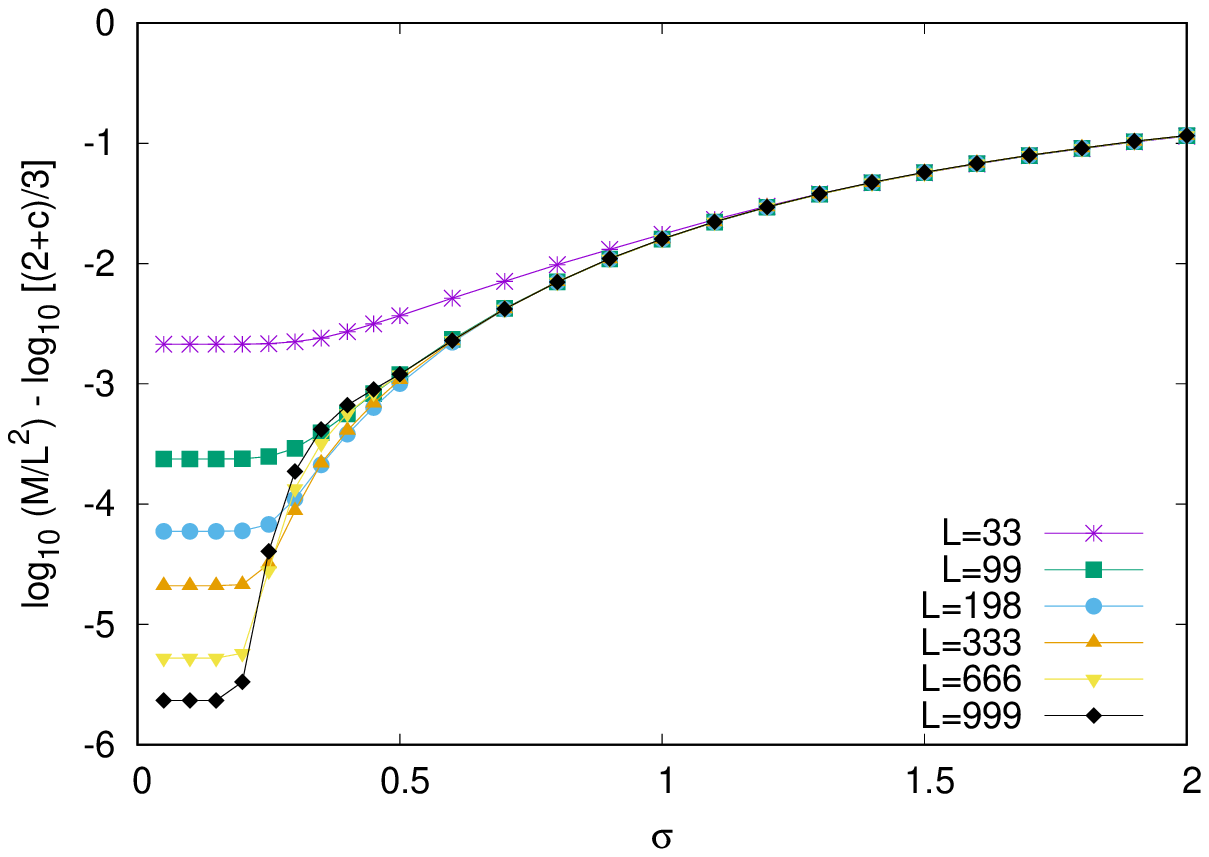} \caption{ 
Metastable states vs. $\sigma$ on a partially diluted triangular 
lattice with $c=0.60$ in the range $0.1 \le \sigma \le2.0$.}
\label{fig8}\end{figure}

\end{document}